\documentclass[]{spie}  
\usepackage[]{graphicx}
\usepackage{fixltx2e}
\usepackage{amsmath}
\usepackage{bbold}

\title{M\&m's:\\ An error budget and performance simulator code for polarimetric systems} 

\author{Maria de Juan Ovelar, Frans Snik, Christoph U. Keller
\skiplinehalf
\small{Sterrekundig Instituut Utrecht, Princetonplein 5, 3584 CC Utrecht, the Netherlands} \\}
\authorinfo{Further author information: (Send correspondence to M.J.O.)\\ E-mail: {\tt m.dejuanovelar@uu.nl}}

\begin{document} 
  \maketitle 

\begin{abstract}

Although different approaches to model a polarimeter's accuracy have been described before, a complete error budgeting tool for polarimetric systems has not been yet developed. 
Based on the framework introduced by Keller \& Snik, in 2009, we have developed the {M\&m's} code as a first attempt to obtain a generic tool to model the performance and accuracy of a given polarimeter, including all the potential error contributions and their dependencies on physical parameters. 
The main goal of the code is to provide insight on the combined influence of many polarization errors on the accuracy of any polarimetric instrument.
In this work we present the mathematics and physics based on which the code is developed as well as its general structure and operational scheme. Discussion of the advantages of the {M\&m's} approach to error budgeting and polarimetric performance simulation is carried out and a brief outlook of further development of the code is also given.

\end{abstract}


\keywords{polarimetry, error budgeting, systems engineering}


\section{INTRODUCTION} \label{sec: }

Polarimetry is a very valuable remote-sensing technique that often yields information that is unobtainable through other techniques and is used in many different fields such as astronomy, Earth observation, biomedical diagnosis or land-mine and target detection, among others.

Nevertheless, the construction of such instruments has often relied on the knowledge obtained through experience than on a formal systems engineering approach, which is common practice when designing optical (imaging or spectroscopic) systems. 
Also, quite often, the polarimetry is implemented as an add on to an existing system which makes it sub-optimal by definition.
It is only now, due to the increasing size and complexity of current instrumentation projects, that this is starting to be demanded. 
For this we need to be able to predict the behavior and accuracy of the polarimetric system in order to optimize the design process.
Only with the implementation of polarimetric error budgeting can the polarimetric performance be fully traded off against optical performance and other merit functions (like cost).

In the systems engineering approach, the design process starts by setting the scientific requirements of the instrument which then must be translated into accuracy and sensitivity requirements. 
Polarimetric sensitivity is defined as the smallest signal that the instrument can detect above the noise and polarimetric accuracy is the uncertainty with which the instrument is able to measure a polarization signal after it has been detected with sufficient S/N\cite{SnikKeller2011}.
The former is limited by (photon) noise and spurious polarization effects that can be created by e.g.~variable atmospheric properties or source variability. 
These effects are to a large extent decoupled from the polarimetric accuracy, and formalisms and codes exist to model them\cite{CassinideWijn2011}.
The accuracy is limited by ``real'' polarization errors, which can be described by Mueller matrices.
The goal of the design process then is to assure that the instrument will indeed meet the accuracy requirements when operating.

With this aim, one needs to simulate the performance of the different preliminary designs and make an estimation of the error budget throughout the system to detect which elements contribute, and in which way, to the final response of the instrument. 
This allows to deal with the limitations of the system in an early design stage.
In optical systems dealing only with intensity it is common practice to make use of error budgeting and performance simulation tools that can carry out such an optimization process. 
Often this optical error budgeting involves adding wavefront errors (which are small, independent and scalar) in an RSS (root sum square) fashion, and force the total error to be smaller than a certain required value.
The individual errors can then be distributed top-down, or added up bottom-up by adopting measured or modeled values, or a combination of both.
So far, to our knowledge, a complete polarimetric error budgeting tool, working with a library of all possible error sources and their possible interactions has not yet been developed for polarimetric systems and this is mainly due to the complexity of the error propagation in these systems.
Tyo (2002)\cite{ScottTyo2002} and Boger et al.~(2003)\cite{Bogeretal2003} introduced polarimetric error budgeting with similar mathematical formalisms as presented in this paper, but their scopes focused on specific polarimetric elements.

Errors in polarimetry have to be expressed as vectors and their values are often larger than the measured signal itself, due to the fact that some elements, essential to perform polarimetry, affect the polarization state of the incoming light in a major way. 
Often the degree of polarization of the signals to be measured is very low, and possibly much lower than the instrumental polarization. 
The main implication of this is that the common algorithms applied to optical systems for error propagation as RSS cannot be applied as such to polarimetric systems. 
In 2009 Keller \& Snik analyzed this problem and developed a mathematical framework to transform the errors into additive ones to make them suitable for error budgeting and estimate the contribution of each physical parameter to the overall matrix\cite{KellerSnik2009}.

Based on this framework we have developed the {M\&m's} code that computes the error propagation through the polarimetric system of all the potential error contributions, e.g.~misalignment and varying material properties, and their dependencies on (global) physical parameters, e.g.~wavelength and temperature.
The ultimate goal of the code is to provide insight of the contribution of error sources to the final polarimetric performance and estimate the polarimetric accuracy of a given design.
The code only pertains to predicting the polarimetric accuracy of a certain instrument as it relates the incoming Stokes vector to the measurement result. 

In Section \ref{sec:math} a description of the mathematical framework and the computational procedure is given. In Section \ref{sec:physics} the physics for the modeling of error sources that M\&m's uses in its library are explained. In Section \ref{sec:codeoverview} a description of the structure and operational modes of the code is given. Section \ref{sec:discussionandoutlook} discusses the advantages and disadvantages of this approach for error budgeting of polarimetric systems and the next steps to be taken in the development of the simulator.

\section{Error propagation in polarimetric systems: the math} \label{sec:math}

\subsection{Mathematical approach} \label{ss:mathapproach}

In the Stokes formalism, every element in an optical system can be described as a $4 \times 4$ matrix called Mueller matrix ($\mathbf{M_{element}}$).
The Stokes vector going out of each element $\mathbf{S_{out}} \equiv (I_{out},Q_{out},U_{out},V_{out})^T$ is then obtained by multiplying this matrix by the incoming Stokes vector $\mathbf{S_{in}} \equiv (I_{in},Q_{in},U_{in},V_{in})^T$, viz.~:

\begin{equation} \label{eq:muellermatrix}
\mathbf{S_{out}} = \mathbf{M_{element}} \mathbf{S_{in}}\,.
\end{equation}

Each element of this matrix represents a relation between the components of $\mathbf{S_{in}}$ and $\mathbf{S_{out}}$ :

\begin{equation} \label{eq:muellermeaning}
\mathbf{M_{element}} =
\left(
\begin{matrix} 
 I_{in} \rightarrow I_{out} & Q_{in} \rightarrow I_{out} & U_{in} \rightarrow I_{out} & V_{in} \rightarrow I_{out}\\
 I_{in} \rightarrow Q_{out} & Q_{in} \rightarrow Q_{out} & U_{in} \rightarrow Q_{out} & V_{in} \rightarrow Q_{out}\\
 I_{in} \rightarrow U_{out} & Q_{in} \rightarrow U_{out} & U_{in} \rightarrow U_{out} & V_{in} \rightarrow U_{out}\\
 I_{in} \rightarrow V_{out} & Q_{in} \rightarrow V_{out} & U_{in} \rightarrow V_{out} & V_{in} \rightarrow V_{out} 
\end{matrix}
\right)\,.
\end{equation}

If we have a set of $n$ elements forming an optical system we then have to multiply the respective element matrices from the last to the first optical element on the light path to get the total matrix for the system $\mathbf{M_{tot}}$.

\begin{equation} \label{eq:totalmatrixdef}
\mathbf{M_{tot}} = \mathbf{M_n} \mathbf{M_{n-1}} \dots \mathbf{M_2} \mathbf{M_1}\,.
\end{equation}

Which obviously relates the incoming and outgoing Stokes vectors as follows:

\begin{equation} \label{eq:totalmatrix}
\mathbf{S_{out}} = \mathbf{M_{tot}} \mathbf{S_{in}}\,.
\end{equation}

Each element's performance, and hence its corresponding Mueller matrix ($\mathbf{M_z}$  with $z = 1, 2, \dots, n$ ), will depend on a certain number ($j_{max}(z)$) of physical parameters each of which will have a particular distribution of values, i.e.~an uncertainty, or fixed offset. 
These uncertainties on each parameter's value will have an effect on the Mueller matrix of the element that can be expressed by the product of the error on the parameter ($\delta p_{j,z}$) and a matrix that tells us how this parameter affects each element of the main matrix, referred to as \textit{``weight"} matrices ($\mathbf{m_{j,z}}$) in the following. 
In this way we can express the Mueller matrix of any element as the sum of a \textit{``main"} matrix ($\mathbf{M}$) and $j_{max}(z)$ \textit{``error"} matrices ($\delta p_j \cdot \mathbf{m_j}$):

\begin{equation} \label{eq:elemsMnms}
\mathbf{M_z}\left(p_1+\delta p_1,\dots, p_j+\delta p_{j_{max}(z)} \right) =
\mathbf{M}(p_1,\dots, p_j) + \delta p_1 \cdot \mathbf{m_1} + \dots +\delta p_j \cdot \mathbf{m_j}+ \dots +\delta p_{j_{max}(z)} \cdot \mathbf{m_{j_{max}(z)}}\,. 
\end{equation}

Now the error matrices, ($\delta p_{j,z} \cdot \mathbf{m_{j,z}}$), can be described by the first order of the Taylor approximation of $\mathbf{M}$ with respect to each parameter $p_{j,z}$ as showed by Keller \& Snik (2009) and Tyo (2002)\cite{KellerSnik2009, ScottTyo2002}. 
This is only valid if (1) the errors are small, compared to the value of the corresponding main matrix value, and  (2) the errors are independent from each other. 
In some cases the error in certain parameters can be such that the first order of the Taylor expansion will not suffice to properly approximate the contribution to the main matrix and that higher orders have to be considered, but as a first step we will use the simplified expression.

Now, to find the total Mueller matrix of a particular system composed of $n$ elements we would have to, following the Stokes formalism, multiply the corresponding Mueller matrices:

\begin{equation} \label{eq:totalMatrixextended}
\mathbf{M_{tot}} = \prod_{z=n}^{1} \left[ \mathbf{M_{z}}\left(p_{1,z},\dots, p_{j_{max}(z),z} \right) + 
\delta p_{1,z} \cdot \mathbf{m_{1,z}} +
\dots +\delta p_{j_{max}(z),z} \cdot \mathbf {m_{j_{max}(z),z}} \right]\,.
\end{equation}

Taking into account the assumptions (1) and (2) made above, the resulting expression can be simplified as follows:

\begin{equation} \label{eq:totalMatrix}
\mathbf{M_{tot}} \approx 
\prod_{z=n}^{1} \mathbf{M_{z}} +
\sum_{z=n}^{1} \sum_{j=1}^{j_{max}(z)} \delta p_{j,z} \cdot \left( \prod_{k=1}^{z+1} \mathbf{M_{k}}\right) \mathbf{m_{j,z}} \left( \prod_{k=z-1}^{k=n} \mathbf{M_{k}}\right)\,.
\end{equation}

\subsection{Implementation of Polarimetric Modulation} \label{ss:compproc}

\begin{figure} [h]
  \centering
   \includegraphics[width=0.55\textwidth]{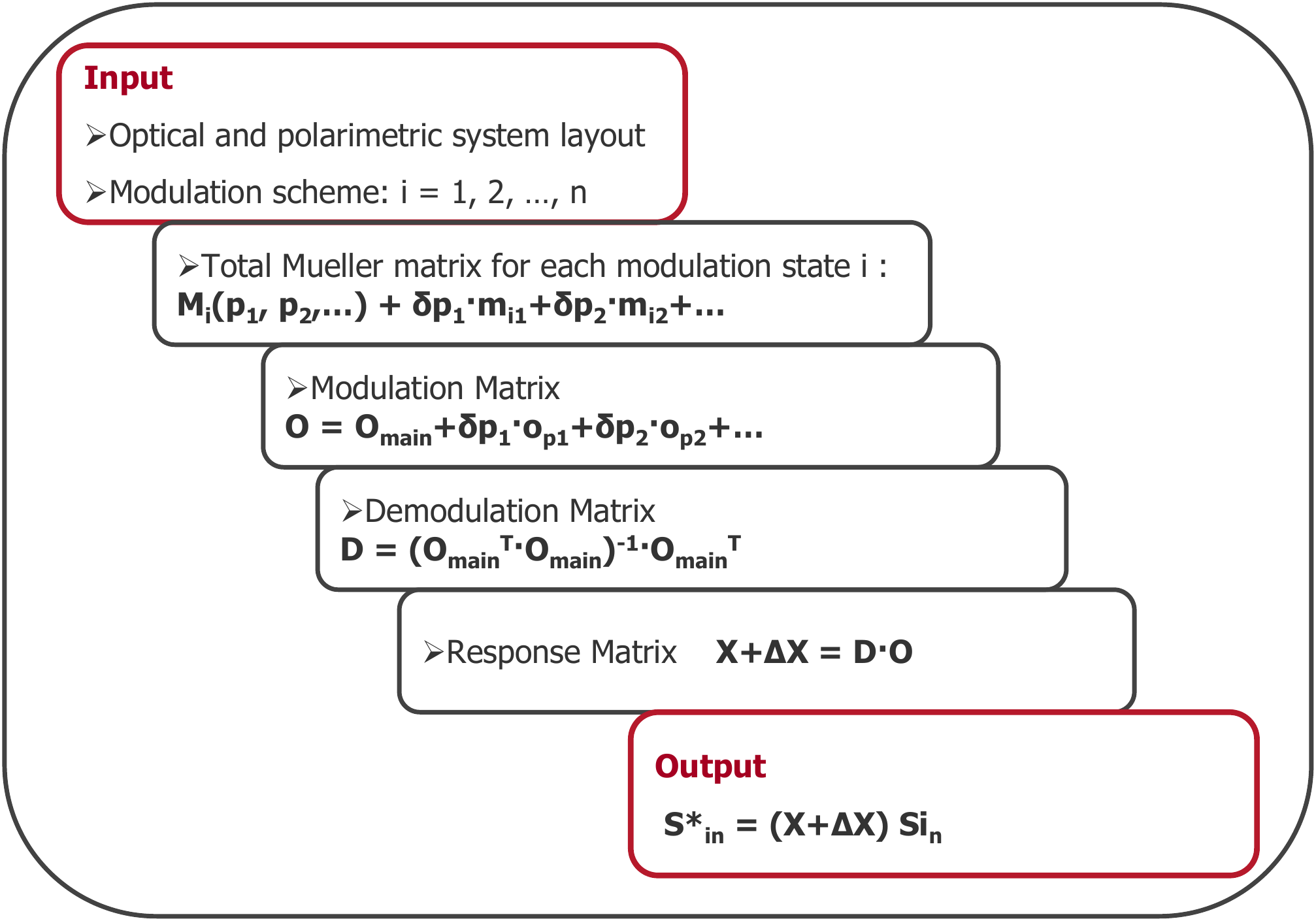}
   \caption{Schematic view of the M\&m's computational procedure. The program starts with the input given in the master script. Then, making use of the M\&m's library, it computes the Mueller matrices for the \textit{n} elements and the final Mueller matrix of the system, for each modulation state \textit{i}. Next step is to build the modulation matrix taking the first row of each main and weight matrices to propagate the error on each parameter into the modulation process. Then the modulation matrix \textbf{O} is split into a main matrix $\mathbf{O_{main}}$ and the corresponding matrices $\mathbf{o_{j,z}}$ for each parameter of the system. The demodulation matrix is either obtained computing the pseudo-inverse of $\mathbf{O_{main}}$ or specified by the user. The response matrix of the polarimeter \textbf{X} is then obtained as the product of the modulation and demodulation matrices and gives the relationship between the measured and the incoming Stokes vectors. The matrix $\mathbf{\Delta X}$ can then be used to perform the error budgeting\cite{Ichimoto2008, ScottTyo2002}.}
   \label{fig:Mnmsprocess}
\end{figure}

The {M\&m's} code is designed to simulate the polarimetric measurement process from calculating the Mueller matrix of a system to giving the final accuracy with which a certain Stokes parameter is measured. The procedure used to simulate the polarimetric measurement is schematically shown in Figure \ref{fig:Mnmsprocess} and detailed below.

Once we have the expression describing how the errors $\delta p_{j,z}$ affect the main Mueller matrix of the system, we can calculate the total Mueller matrix for each modulation state (\textit{i}). 
This enables us to build the modulation ($\mathbf{O}$) and demodulation ($\mathbf{D}$) matrices, including the error components, and eventually propagate the error to the final output of the polarimetric measurement process. 
The type of modulations implemented on the code are only spatial modulation, temporal modulation or a combination of both. 
Spectral modulation or channelled spectropolarimetry has a demodulation procedure based on Fourier transforms and therefore can't be represented by a linear operation.

Continuing with the nomenclature presented in Section \ref{ss:mathapproach}, let us consider an optical system composed of $n$ elements, each of which depends on a certain, and generally different, number of parameters ($p_{j,z}$, with $j = 1, 2, \dots, j_{max}(z)$ and $z = 1, 2,\dots, n$) and a modulation scheme composed of $m$ modulation states. 
Note that we will then have an $h = \sum_{z = 1}^{n} j_{max}(z)$ total number of parameters for the system, and $j = 1, 2, \dots, j_{max}(z)$ inside each $z$ optical element. Table \ref{tab:indices} shows, for clarity, the definition and ranges of the indexes we will use in the following.

\begin{table}[h]
\caption{Definition of indexes}
\begin{center}
\begin{tabular}{|l|c|l|}
\hline
Optical element & $z$ & $z = 1, 2, \dots, n$ \\ \hline
Parameter in $z$ & $j$ & $j = 0, 1, \dots, j_{max}(z)$ \\ \hline
Modulation state & $i$ & $i = 1, 2, \dots, m$ \\
\hline
\end{tabular}
\end{center}
\label{tab:indices}
\end{table}

For each modulation state $i$ we would have a total Mueller matrix for the system:

\[i = 1,\quad  \mathbf{M_{tot,1}} = \mathbf{M_1} + \delta p_{1,1} \cdot \mathbf{m_{1,1,1}} + \dots + \delta p_{j,z} \cdot \mathbf{m_{j,z,1}} + \dots + \delta p_{j_{max}(n),n} \cdot \mathbf{m_{j_{max}(n),n,1}}\] 

\[i = 2,\quad \mathbf{M_{tot,2}} = \mathbf{M_2} + \delta p_{1,1} \cdot \mathbf{m_{1,1,2}} + \dots + \delta p_{j,z} \cdot \mathbf{m_{j,z,2}} + \dots + \delta p_{j_{max}(n),n} \cdot \mathbf{m_{j_{max}(n),n,2}}\]

\[\vdots\]

\[i = m,\quad \mathbf{M_{tot,m}} = \mathbf{M_m} + \delta p_{1,1}\cdot \mathbf{m_{1,1,m}} + \dots + \delta p_{j,z}\cdot \mathbf{m_{j,z,m}} + \dots + \delta p_{j_{max}(n),n} \cdot \mathbf{m_{j_{max}(n),n,m}}\] \,.

Each $\mathbf{M_{tot,i}}$ represents the behavior of the instrument for each modulation state. 
There will be only a few Mueller matrices that will change with ($i$), since is only the modulator (spatial or temporal) that changes state. The code will be able to distinguish between those ones and the ``static" ones so the calculation of the total Mueller matrix can be shortened.

Now if we have an incoming Stokes vector $(\mathbf{S_{in}})$ going through the system, for each modulation state we will get an intensity $(I_i)$ measured at the detector. 
These intensities form a vector at the end of the modulation cycle, $\mathbf{I_{meas}} \equiv (I_1, I_2, \dots, I_m)$. 
A linear combination of these $m$ intensities will allow us to infer the state of polarization of the incoming light. 
This linear relation between the $m$ intensities measured by the detector and the incoming stokes vector is given by the so-called modulation matrix $(\mathbf{O})$,

\begin{equation} \label{eq:modulation}
\mathbf{I_{meas}} = \mathbf{O} \mathbf{S_{in}}\,.
\end{equation}

The modulation matrix $\mathbf{O}$ is an $m \times 4$ matrix built by compiling the first row of every $\mathbf{M_{tot,i}}$ matrix. 
This can be understood remembering that it is in this row of the Mueller matrix where we find how the components of the incoming Stokes vector get transformed into output intensity Eq. \ref{eq:muellermeaning}. 
Given the structure that these Mueller matrices now have, i.e.~being split in a sum of matrices, we will obtain a modulation matrix that is also a sum of matrices. 
We will get a \textit{main} modulation matrix ($\mathbf{O_{main}}$) obtained by compiling the first rows of main matrices $\mathbf{M_i}$ and  $h$ matrices $\mathbf{o_{j,z}}$, obtained by compiling the first rows of weight matrices $\mathbf{m_{j,z,i}}$, each one associated to one physical parameter of the system. 

\begin{equation} \label{eq:Ostructure}
\mathbf{O} = \mathbf{O_{main}} + \delta p_{1,1} \cdot \mathbf{o_{1,1}} + \dots + \delta p_{j,z} \cdot \mathbf{o_{j,z}} + \dots + \delta p_{j_{max}(n),n} \cdot \mathbf{o_{j_{max}(n),n}}\,.
\end{equation}

$\mathbf{O_{main}}$ tells us how the intensity is measured without taking into account the deviations of the parameters. 
The "separated" propagation of the errors to this point allows us to analyze how they affect the intensity measurement directly, so we can easily detect the elements and parameters that have more impact on the measurement before the demodulation process.

It is at this point of the process when we can implement detector errors such as bias drift and nonlinearities (see Section \ref{sec:physics} for further description).

Once we have the measured intensity vector the demodulation process starts. The basic goal of this process is to infer $\mathbf{S_{in}}$ from the set of intensity measurements $\mathbf{I_{meas}}$. 
It is clear from Eq.\ref{eq:modulation} that this can be done by the inverted matrix of $\mathbf{O}$, which we will refer to as $\mathbf{D}$, the \textit{demodulation} matrix. 
This is a trivial mathematical problem only if $\mathbf{O}$ is a square matrix, which is the case for modulation schemes composed of four measurements. 
If the modulation scheme comprises a larger number of modulation states, i.e. $m > 4 $, the solution for the inversion of $\mathbf{O}$ is no longer unique and we have an infinite number of \textit{demodulation} matrices that fulfill the condition $\mathbf{D} \mathbf{O} = \mathbf{1}$. 
Then the new problem is to find the demodulation matrix that optimizes the efficiencies of the system\cite{delToroIniestaCollados2000, ScottTyo2002}. In 2000, del Toro Iniesta \& Collados\cite{delToroIniestaCollados2000} showed that the solution for this optimization problem is found by means of the Moore-Penrose pseudo-inverse matrix:

\begin{equation} \label{eq:MoorePenrose}
\mathbf{D} = (\mathbf{O^T}  \mathbf{O})^{-1} \mathbf{O^T}\,.
\end{equation}

At this point we need to make some considerations.
Given expressions (\ref{eq:Ostructure}) and (\ref{eq:MoorePenrose}) we would now have to calculate the Moore-Penrose pseudo-inverse matrix of a sum of a large number of matrices, depending on the total number of parameters in our system. 
Furthermore, we want to keep the errors $\delta p_{j,z}$ as unknown variables up to the end of the process so we can analyze the system in a generic way, without considering specific values for the errors. 
There are various mathematical formulae and theorems to obtain the inverse of a sum of matrices but , as far as we know, there is no way to obtain analytically the inverse of a structure like this particular one, specially with a variable number of matrices. 
The problem of how errors propagate through the non-linear process of matrix inversion has been analyzed by Asensio Ramos \& Collados in 2008 for gaussian errors\cite{AsensioRamosCollados2008} but systematic errors have a complete different distribution which makes them unsuitable for this procedure. 
A simplified approach is to obtain $\mathbf{D}$ by means of an ``ideal" or ``best known" modulation matrix, which would be $\mathbf{O_{main}}$ in our case.

This can be justified arguing that during data reduction one has to choose a demodulation matrix. 
The best knowledge one has of the modulation matrix is obtained after modeling (or calibrating). This matrix would be $\mathbf{O_{main}}$ for us, since it represents de \textit{``ideal"} behavior of the modulation matrix.
In this way we can also propagate the error in the parameters as unknown variables through the demodulation process. 

Therefore:

\begin{equation} \label{eq:MoorePenroseOmain}
\mathbf{D^*} = (\mathbf{O_{main}^T}  \mathbf{O_{main}})^{-1} \mathbf{O_{main}^T}\,.
\end{equation}

There is also the possibility of using a pre-defined demodulation matrix which, even though it may not be the optimal one, will be required in many practical situations.
This can be easily implemented in the code by providing the demodulation matrix together with the input and not deriving it from the calculated modulation matrix. 
Also the code needs to be suitable for partial polarimeters, i.e.~polarimeters only measuring a part of the Stokes vector, for which the modulation matrix when written as a $m \times 4$ matrix is singular and non-invertible. Therefore the code would have to be able to reduce the dimensionality.

Once we have the demodulation matrix we can finally find the vector that will represent the ``measured" incoming Stokes vector, $\mathbf{S^*_{in}}$,

\begin{equation} \label{eq:demodulation}
\mathbf{S^*_{in}} = \mathbf{D}  \mathbf{I_{meas}}\,.
\end{equation}

Note that this vector is not a Stokes vector since is obtained through the matrices $\mathbf{O}$ and $\mathbf{D}$ which are not Mueller matrices.

Ideally this output will be exactly the Stokes vector that comes in the polarimeter so that:

\begin{equation} \label{eq:idealpol}
\mathbf{S^*_{in}} = \mathbb{1} \mathbf{S_{in}}\,.
\end{equation}

Of course this is not the case since no instrument is exempt from creating some uncalibrated instrumental polarization and cross-talk along the measurement process. 
To take this into account it is useful to introduce the concept of a \textit{response matrix} $\mathbf{X}$ \cite{Ichimoto2008}. 
This matrix, which is not a Mueller matrix, has dimensions of $4 \times 4$  and can be defined as the one representing the measurement process i.e.~it relates the incoming Stokes vector ($\mathbf{S_{in}}$) with the vector $\mathbf{S^*_{in}}$ that we obtain in the end as the ``measured" Stokes vector

\begin{equation} \label{eq:response}
\mathbf{S^*_{in}} = \mathbf{X} \mathbf{S_{in}}\,.
\end{equation}

Again, in ideal conditions $\mathbf{X}$ would be the identity matrix, but since we have errors in the process it will have values different from unity. 
In the end, what we aim to do with the M\&m's code is to model this response matrix, so find the range of values its elements can take on, and this is done by propagating our error-dependent matrices $\mathbf{o_{j,n}}$ up to the end of the inversion process. 
Considering equations \ref{eq:modulation}, \ref{eq:demodulation} and \ref{eq:response} it is obvious that we can obtain $\mathbf{X}$ by means of $\mathbf{D}$ and $\mathbf{O}$ as follows:

\begin{equation} \label{eq:responsedef}
\mathbf{X} = \mathbf{D} \mathbf{O}\,. 
\end{equation}

If we now introduce the expression for $\mathbf{O}$, Eq. \ref{eq:responsedef} becomes:

\begin{equation} \label{eq:responsedevel}
\mathbf{X} = \mathbf{D} [ \mathbf{O_{main}} + \delta p_{1,1} \cdot \mathbf{o_{1,1}} + \dots +\delta p_{j,z} \cdot \mathbf{o_{j,z}} + \dots + \delta p_{j_{max}(n),n} \cdot \mathbf{o_{j_{max}(n),n}}]\,.
\end{equation}

which applying the distributive property gives:

\begin{equation} \label{eq:responsedevel2}
\mathbf{X} = \mathbf{D} \mathbf{O_{main}} + \delta p_{1,1} \cdot (\mathbf{D} \mathbf{o_{1,1}}) + \dots +\delta p_{j,z} \cdot (\mathbf{D} \mathbf{o_{j,z}}) + \dots + \delta p_{j_{max}(n),n} \cdot (\mathbf{D} \mathbf{o_{j_{max}(n),n}})\,.
\end{equation}

If the demodulation matrix is the pseudo-inverse of the main modulation matrix, i.e.~we have computed $\mathbf{D}$ instead of using a pre-defined one, the first term of the right side of the equation is the identity matrix and each $(\mathbf{D} \mathbf{o_{j,z}})$ term represents a variation of the response matrix depending on the corresponding $p_{j,z}$. 

\begin{equation} \label{eq:responsefin}
\mathbf{X} = \mathbb{1} + \delta p_{1,1} \cdot \mathbf{\Delta X_{1,1}} + \dots +\delta p_{j,z} \cdot \mathbf{\Delta X_{j,z}} + \dots + \delta p_{j_{max}(n),n} \cdot \mathbf{\Delta X_{j_{max}(n),n}} = \mathbb{1} + \mathbf{\Delta X}
\end{equation}

In this way, we have propagated the errors up to the final matrix that models the complete measurement process.

\subsection{Error budgeting} \label{ss:errorbudgeting}

The matrices $\mathbf{\Delta X_{j,z}}$ and $\mathbf{\Delta X}$, from Equation \ref{eq:responsefin}, can be seen as \textit{``accuracy matrices"} to be directly used for error budgeting by comparing them with the requirements set for the polarimeter. 

This can be done in different ways such as comparing the complete accuracy matrix ($\mathbf{\Delta X}$) with a ``requirements matrix" that one builds from the scientific requirements\cite{Ichimoto2008} (Eq. \ref{eq:errorbudgetIchimoto}):

\begin{equation} \label{eq:errorbudgetIchimoto}
\mathbf{\Delta X} \leq \mathbf{\Delta X_{req}}\,.
\end{equation}

To simplify the process only the most stringent components of the $\mathbf{\Delta X_{req}}$ can be considered.

Another approach is to assume an input Stokes vector ($\mathbf{S_{in,def}}$) and compute the difference or vector norm between the this input and the measured Stokes vectors ($\mathbf{S^*_{in}}$) obtained by multiplying with $\mathbf{\Delta X}$ (Eq. \ref{eq:errorbudgetTyo}) \cite{ScottTyo2002} :

\begin{equation} \label{eq:errorbudgetTyo}
\sum_{k = 1}^{4} \left(\mathbf{\Delta X} \cdot \mathbf{S_{in,def}}\right)_{k}^2\leq ||\epsilon||^2 \,.
\end{equation}

The requirement on polarimetric accuracy is then described by the vector norm $||\epsilon||$, and should be fulfilled for all possible input Stokes vector.

\subsection{Implementation of Calibration} \label{ss:calibration}

The code can be set up such that it describes the full procedure of calibration, after the Mueller matrices for the calibration optics including error terms have been introduced.
These results should then be fed back into the code to describe the calibrated polarimetric measurements.
However, the error propagation from the calibration to the measurement is not straightforward and has only be mathematically described in the case of Gaussian errors\cite{AsensioRamosCollados2008}.
Possibly a complete Monte Carlo simulation has to be performed to fully model this process.

In any case, real calibration results and the estimated errors thereupon can be implemented directly by replacing the pertinent part of the modeled Mueller matrix train.

\section{Overview of polarimetric errors: the physics} \label{sec:physics}

Now that we have defined the mathematical procedure we need to introduce the physics behind the error modeling, i.e.~determine $\mathbf{m_{j,z}}$.
An error budgeting code like {M\&m's} is only useful once it is able to consider almost any kind of error to any kind of optical element. 
Based on the current literature, an extensive library is therefore created out of which most relevant Mueller matrices and error matrices can be taken or computed.

Errors affecting polarimetric measurements can be widely ranging in both type and scale and can arise from many different sources like missalignements, manufacturing errors or variations of the working temperature. 
In polarimetry, every element before the polarimeter, even a simple glass plate, can affect the measurement. Therefore a careful analysis of each element must be done in order to model its behavior as accurately as possible.
 
There are two main ways in which this can be done. 
One way is based on the direct characterization of the elements in the lab, or what could be seen as the empirical approach. 
This method, of course, provides a quite complete knowledge of the particular element one is considering but lacks of potential for generalization and it is limited by the accuracy of the measuring system. 
Another approach is to start from the general analytical expression of the element's Mueller matrix depending on various physical parameters and derive the error matrices as a first order Taylor expansion to each parameter.
In our case we consider both possibilities, so the code can estimate an approximation of the \textit{weight matrices} for each parameter and it can also include error matrices that are already determined.

\subsection{Parameters} \label{ss:parameters} 

Parameters can be separated in two groups, the ones affecting the Mueller matrix of only one particular element, e.g.~the birefringence of the material of a retarder or the extinction ratio of a polarizer, and those which are common for all the system, e.g.~wavelength and temperature. The latter will be represented in the code by global parameters. 
This means that all Mueller matrices inasmuch they depend on wavelength and temperature will automatically have error matrices associated to them that scale with wavelength or temperature.
Thus the variation of system performance with wavelength and temperature can be estimated from the code's output.
It is clear that with only a first-order approximation of Mueller matrix variation with wavelength, only relatively narrow band-widths can be considered.
If one element has a different temperature than the others, its Mueller matrix at that temperature and corresponding error matrix for temperature changes can be also inserted separately.

\subsection{Library of Mueller Matrices and Error Matrices} \label{ss:errors} 

The following describes the implementation of Mueller matrices of common optical elements.
It also describes which error sources can be used in the code and under which assumptions.

\subsubsection{Rotations}
This is a special case because it is not an element per se. 
A single rotation Mueller matrix describes the rotation of the $[Q,U]$ coordinate system. 
The only error in this process is an offset to this rotation (or a certain distribution of offsets).
For the case of a freely rotating element, two rotation matrices need to be inserted, one before and one after the element (which can be a compound of several Mueller and error matrices).
The amount of rotation and the error thereof is identical, but have an opposite handedness, unless the elements in between modify Stokes coordinate system like an odd number of mirrors do.

Rotation is often used in temporal modulation, so the variation of the Mueller matrix with modulation state is generically implemented in this function.
 
\subsubsection{Linear Polarizers}

The ideal Mueller matrix of a linear polarizer assumes that linear polarization in the $+Q$ direction gets transmitted.
In the case of spatial modulation by using a polarizing beam-splitter, this direction can be switched to $-Q$ in the input script that handles the modulation sequence.

The following error terms are implemented:
\begin{itemize}
\item Limited extinction ratio. The extinction ratio is here defined such that its value is $<$1. The error matrix for limited extinction ratio is described by Snik (2006)\cite{Snik2006}. This error necessarily has a first and second order.
\item Depolarization cf.~Nee et al.~(1998)\cite{Nee1998}. 
\item Field-of-view effects. This is described with the 3D geometry of the polarizers axes compared to the input polarization.
\end{itemize}

\subsubsection{Retarders}

The most frequently used retarders are based on birefringent (liquid) crystals.
The retardance can be computed once the material's birefringence and the plate's thickness is known: 

\begin{equation}
\delta = 2\pi \frac{(n_e - n_o) \cdot d}{\lambda}\,.
\end{equation}

The refractive indices for many crystals are listed by Ghosh (1998)\cite{Ghosh}.

Various error sources are readily derived from this:
\begin{itemize}
\item Variation with wavelength.
\item Variation of the thickness.
\item Offset of the birefringence.
\item The variation of the birefringence with temperature can be computed from thermo-optic coefficients listed by Ghosh (1998)\cite{Ghosh}. See Hale \& Day (1988)\cite{HaleDay}. Also the thermal expansion of the plate is taken into account.
\item The variation of retardance upon a change of incidence angle and azimuth, as described by the equations in Evans (1949)\cite{Evans}.
\item Dichroism, as determined by the Fresnel equations for the two polarization directions along the two crystal axes.
\item Polarized fringes due to multiple reflections within the birefringent plate manifest themselves as a sinusoidal partial polarization pattern as a function of wavelength, as well as a wavelength-dependent modification of the retardance. These effects are calculated by using the Berreman calculus as described by Weenink et al. (2011)\cite{Weenink}. The mitigation of these effects by using AR coatings can be implemented ad-hoc.
\end{itemize}

Since most wave-plates used for instrumentation are actually compounds of two or more layers of different materials, it makes sense then to consider compound wave plates as unit elements in the system.
Several frequently used wave plates will be predefined in the {M\&m's} library: achromatic and superachromatic wave plates and liquid crystals like Liquid Crystal Variable Retarders (LCVRs) en Ferroelectric Liquid Crystals (FLCs).
For the first cases, the alignment between the two plates becomes an important error parameter.
For the liquid crystals, the switching angles (that often drive the modulation sequence) will be subject to errors.

Also a standard Fresnel rhomb will be implemented. The main error sources there will be:
\begin{itemize}
\item Variation of retardance with incidence angle and azimuth.
\item Stress birefringence.
\end{itemize}

\subsubsection{Mirrors}

The Mueller matrix of a mirror is based on the models applied to accurate ellipsometry performed by Van Harten et al.~(2009)\cite{vanHarten}. 
According to it, a mirror can be characterized knowing the following parameters: the incidence angle ($\alpha$) and wavelength ($\lambda$) of the incoming light, the complex index of refraction of the metal, and, if present, the thickness and index of refraction of the dielectric film layer on top of the mirror's surface. 
This layer can be an artificial overcoating or can occur naturally due to the growth of the aluminum oxide layer\cite{vanHarten}. 
The equations that lead to a mirrors Mueller matrix are numerous and complex and directly describe the dependence of the Mueller matrix on various physical parameters:

\begin{itemize}
\item Variation with wavelength due to the variation of the refractive indices.
\item Variation with incidence angle and azimuth.
\item Aging which involves a growth of the layer of dielectric material on the mirror. Also a dust layer can be described with an effective growth of this layer, see Snik et al.~(2011)\cite{SnikellipsoII}.
\end{itemize}

\subsubsection{Glass Components}
It is often assumed that glass components like lenses do not modify or create polarization, and therefore the ideal Mueller matrix is the unit matrix.
Various error sources can however exist:

\begin{itemize}
\item Glass components can introduce linear polarization if a ray of light hits it at a non-normal incidence. This is fully determined by the Fresnel equations, after linearizing those for the incidence angle. It needs to be computed at every glass-air interface.
\item Stress birefringence occurs in every piece of glass. A certain direction (or distribution) of the stress tensor needs to be assumed.
\end{itemize}

\subsubsection{Detectors}

Although the action of a detector cannot be described by a Mueller matrix, its non-ideal properties can adversely affect the polarimetric accuracy.
Fortunately, these error terms can easily be implemented using a similar formalism.

It is assumed that an ideal detector linearly converts the intensity to a data number.
This corresponds with a fully multiplicative term comparable with a diagonal Mueller matrix.
Two important error terms are:
\begin{itemize}
\item Bias drift or uncorrected stray light.
\item Detector non-linearity. This obviously needs to be described with a second order (or higher) error term.
\end{itemize} 
Keller (1996)\cite{Keller1996} has shown that such errors can couple with instrumental polarization to create spurious signals.

\section{Overview of the code} \label{sec:codeoverview}

The {M\&m's} is a code written in Python which makes it accessible to a broad audience cost-free. The name arises from the basic structure of the Mueller matrices (Eq. \ref{eq:elemsMnms}) that is propagated all along: the ``main" matrix ($\mathbf{M}$) and the ``weight" matrices ($\mathbf{m}$).It is structured in two main parts, the \textit{``input/master script"} and the \textit{``M\&m's library"}.\\

The \textit{input/master script} represents the input to the program. 
In it, the user has to specify all the information about the set up to be simulated, e.g. type and order of the optical elements on the light path, modulation states, global parameters, rotations and misalignments between them, etc....
In this input one can also provide particular matrices to be used instead of those from the library such as specific calibrated error matrices for a certain element or demodulation matrices. 
The user also needs to specify the mode in which the code should operate.

The \textit{M\&m's library} is a compilation of functions that can be separated in two groups according to their functionality. 

The first group is composed of \textit{``element functions"} computing the Mueller matrix of typical optical elements in terms of the physical parameters in which the element depends on and deriving the corresponding weight matrices. 
The library contains functions to model mirrors, glass elements, waveplates and polarizers among others. 

\begin{figure} [h]
  \centering
   \includegraphics[width=0.80\textwidth]{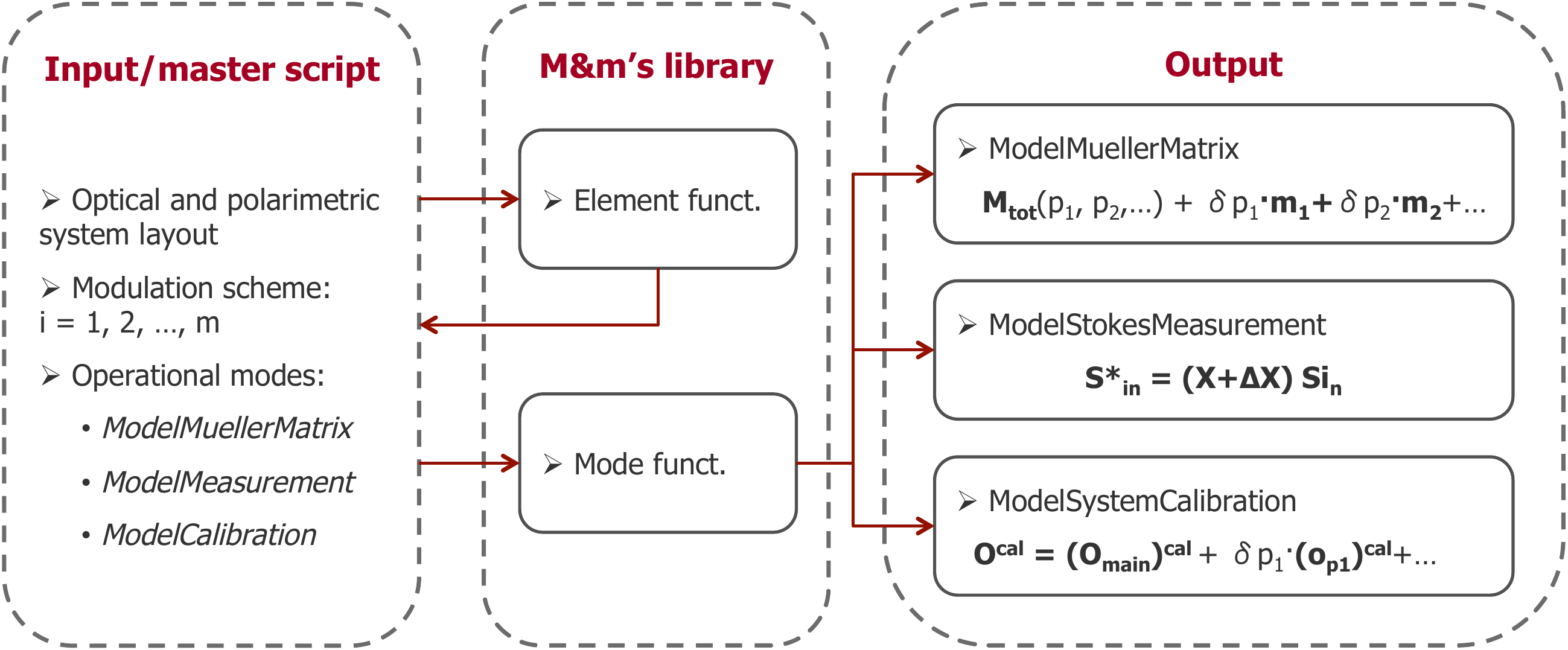}
   \caption{Flow diagram of the M\&m's operation. The master script sends the input information to the library's selected element functions to be used. The output of this operation is sent to the master script which then sends the matrices to the selected mode function producing the final required output. }
   \label{fig:OverviewProcess}
\end{figure}

The second group is formed by the \textit{``mode functions"} which, depending on the mode selected, operate with the Mueller matrices obtained for the elements in order to get the desired output. Figure \ref{fig:OverviewProcess} shows a flow diagram of the information through the different elements and stages of the code.

With the aim of making it suitable for different types of analysis it operates in three modes:

\begin{itemize}
\item\textit{ModelMuellerMatrix}
\item\textit{ModelStokesMeasurement}
\item\textit{ModelSystemCalibration}
\end{itemize}

The first mode, \textit{ModelMuellerMatrix}, computes the total Mueller matrix of a given system. 
It takes the matrix for each element that come as the output of the \textit{element functions} and then multiplies the matrices following the formalism explained above. It gives the total Mueller matrix for the system as an output with the structure shown in Eq. \ref{eq:totalMatrix}. 
This is useful to analyze how the different element/parameters will affect the final behavior of the polarimetric set up but also to estimate the Mueller matrix of optical systems upstream, i.e.~the telescope in front of the polarimeter. 


The second mode, \textit{ModelStokesMeasurement}, computes the vector that it is measured at the end of the process ($\mathbf{S^*_{out}}$), i.e. it propagates the error matrices through the modulation and demodulation operations. 
In the end we get the correspondence between the incoming and the measured Stokes vector by means of the response matrix \textbf{X} which contains the effect of the physical parameters of the system on the measurement, represented by the propagated \textit{weight matrices} (\textbf{$\Delta$X}).

The third mode, \textit{ModelSystemCalibration} computes the polarimetric modulation matrix as obtained after calibration and the errors pertaining to these results. This mode has not yet been implemented since, as we pointed out before, the error propagation through this operation to a calibrated measurement model needs to be investigated. 

\section{Discussion and outlook} \label{sec:discussionandoutlook}

In this manuscript we have presented the mathematical, physical and operational basis of the {M\&m's} code for simulating the performance of a given polarimetric system. 
The mathematical framework developed by Keller \& Snik allows us to propagate the contribution of the different errors separately up to the end of the measurement process. 
The main advantage of this approach is that it allows us to, at any point of the measurement process, identify the critical sources of deviation in the measurement without having to specify any particular error for the parameters. 
Furthermore the matrix ($\mathbf{\Delta X}$), which is dependent on the $\delta p_j$ scalar errors, can be directly compared to the accuracy matrix required for the system to set the tolerances on the parameters, because we have propagated $\delta p_j$ as unknown variables through the process.

The main disadvantage is that it relies on assumptions (1) and (2) presented in Sec. \ref{ss:mathapproach} (errors are small and independent) when aproximating the weight matrices by the first term of the Taylor expansion of the Mueller matrix. 
Higher order terms may therefore need to be implemented.

The code will be verified by comparing its results with those from a full Monte Carlo simulation (which can be done by using the same library of Mueller matrices), and with the results of a dedicated lab set up.
At the end only a Monte Carlo simulation or lab measurements can fully characterize the system, but the {M\&m's} approach provides a better insight on the system's dependencies. 
This allows us to easily detect, early in the design process, which ones will be the limiting elements/parameters when analyzing the accuracy that different designs, considered for a polarimetric system, can provide.
It also can easily estimate the instrumental polarization introduced by optical systems, keeping also track of which elements have more impact.

The code aims to be open, so it can be used by anyone, general, so it can be applied to any polarimetric design, and complete in terms of error modelling. 
In this sense, we will keep the library open so it can be easily updated with new matrices for elements and errors. 

The code is still in an early stage of development and it will be made accessible to the public as soon as it has been verified, and most relevant errors have been implemented.

The following upgrades to the code are still under consideration:
\begin{itemize}
\item A complete description of a distribution (random or systematic) of the physical parameters that drive the errors. This way, after propagation, the error contributions can be added in an RSS fashion, see Keller \& Snik (2009).
\item Adding the possibility of introducing higher-order error terms.
\item Full implementation of the simulation of the calibration process and the calibrated measurement process. It may be necessary to implement a Monte Carlo simulation of the calibration process to fully propagate the systematic errors, as currently only Gaussian errors can be propagated \cite{AsensioRamosCollados2008} .
\item Integration with the optical design, such that all Mueller matrices are explicitly dependent on the local incidence angles and azimuths. For converging beams, an average Mueller matrix plus a distribution of values around that can thus be obtained. A link with ZEMAX would be a logical choice for this. A simple simple solution would be to parameterize the distance to the pupil plane for each element and supply a range of incidence angle and azimuths.
\end{itemize}


\end{document}